\documentclass[runningheads]{svmult}
\usepackage{makeidx}   
\usepackage{graphicx}  
\usepackage{subeqnar}  
\usepackage{multicol}  
\usepackage{cropmark} 
\usepackage{physprbb}  
\begin{document}
\title*{Self-Interacting Cold Dark Matter Halos}
\toctitle{Self-Interacting Cold Dark Matter Halos}
%
%
\titlerunning{Self-Interacting Cold Dark Matter Halos}
%
\author{Andreas Burkert}
\authorrunning{Andreas Burkert}
%
%
\institute{Max-Planck-Institut f\"ur Astronomie,\\
K\"onigstuhl 17,\\
D-69117 Heidelberg,\\
Germany}

\maketitle              

\begin{abstract}
The evolution of halos consisting of weakly self-interacting dark matter particles
is summarized. The halos initially contain a central density cusp as predicted by
cosmological models. Weak self-interaction leads to the formation 
of an isothermal, low-density core which is in agreement with published data
on the rotation curves of dwarf galaxies. However, subsequently, core collapse
leads to an even steeper central density cusp. In order to explain
the observed dark matter rotation curves by weak interaction, the cross section must be
in a narrow range around $\sigma/m_p \approx 0.1 cm^2 g^{-1}$.
\end{abstract}

\section{Introduction}
Standard cosmological models predict that structure formation in the universe is dominated
by a collisionless, gravitationally interacting, cold dark matter (CDM) component. This theory
has been very successfull in explaining the observed large-scale structures which might
not be very surprising given the fact that large-scale structure formation is dominated 
by the initial power spectrum which to some extent can be fine-tuned.

A more stringent test of the CDM scenario are non-linear dark matter structures,
that is dark matter halos on galactic scales which
 have decoupled from the expansion of the universe
and have achieved a quasi-static virial equilibrium state. The internal structure of 
virialized dark matter halos is a result of the collisionless gravitation interaction
of dark matter particles which go through a phase of violent relaxation during the merging epoch.
Cold dark matter halos therefore provide interesting insight into the nature of dark matter and its interactions
and are to a lesser extent affected by the initial power spectrum.
Cosmological simulations (Navarro et al. 1997) have indeed
shown that virialized CDM halos are reasonably approximated by a universal density distribution, which 
originates from the energy and angular momentum redistribution during the relaxation phase and which
is not sensitive to the chosen initial conditions. However it recently has also become clear that
on these scales the predictions of CDM models are not in good agreement with several observations.

Recent very high resolution simulations (Klypin et al. 1999, Moore et al. 1999) have confirmed the 
earlier claims by Navarro et al. (1997) that CDM halos are singular near their center with a density
profile that diverges as $\rho \sim r^{-\gamma}$, where $\gamma \approx 1.5$. Such cores appear
inconsistent with published data on the rotation curves of galaxies (Moore 1994, Flores \& Primack
1994, Burkert 1995, Salucci \& Burkert 2000) although part of this inconsistency might be due to limitations of the 
data (van den Bosch \& Swaters 2000). 
The calculations have also shown that the predicted number and mass distribution
of galaxies in galactic clusters is consistent with the observations. However, on scales of
the Local Group, roughly one thousand dark halos are expected, whereas less than one hundred satellite
galaxies are observed. This disagreement can also be attributed partly to the high central densities of
dark halos which stabilize them against tidal disruption. Mo et al. (1998) and lateron Navarro
\& Steinmetz (2000) found that CDM models reproduce well the I-band Tully-Fisher slope between
the rotational speed of galactic disks and their luminosity. They fail however to match the zero-point.
Whereas galactic disks with rotation speeds of order 200 km/s
have radial scale lengths of order 3 kpc, the cosmological models predict scale lengths of 300 pc
for the same rotational velocity. Again, this problem can be traced to the excessive central concentrations
of cold dark halos and their stability against tidal disruption which allows satellite galaxies with
an embedded gas component to spiral into large CDM halos by this loosing 90\% of their angular momentum
by dynamical friction before the gas finally decouples from its surrounding dark component and
settles into the equatorial plane.

Motivated by these problems, Spergel \& Steinhardt (2000) proposed a finite cross-section for
elastic collisions for CDM particles, such that their mean free path  is short in halo cores but
long in their outer parts. They suggested that this model could alleviate many of the difficulties,
if the cross section would be of order $\sigma_* = \sigma/m_p \approx 1 cm^2 g^{-1}$.

The Spergel and Steinhard model has motivated a large number of follow-up studies. For example,
Ostriker (2000) demonstrated that weak self-interaction would lead to the growth of massive black
holes in the centers of galactic spheroids through the accretion of dark matter. Miralda-Escud\'{e} (2000)
pointed out that collisional dark matter might produce galaxy cluster which are rounder than observed.
Mo \& Mao (2000) and Firmani et al. (2000) investigated the effect of self-interaction on galactic
rotation curves. Hogan \& Dalcanton (2000) considered how the structural properties of weakly self-interacting
CDM halos scale with their mass.

\section{The Evolution of Weakly Self-Interacting Dark Halos}

Detailed numerical simulations of individual dark matter halos have been presented by several authors.
Moore et al. (2000) and Yoshida et al. (2000) simulated cluster evolution in the fluid limit
of very large interaction cross sections. In this case, collisional dark matter produces even more cuspy
profiles than collisionless CDM, leading to poorer fits to published rotation curves.
Burkert (2000) and lateron Kochanek \& White (2000) studied the secular evolution of virialized 
dark matter halos for more realistic values of the collision cross section using a Monte-Carlo approach.

If the mean free path is not very small compared to the length scale of the system, dark matter 
cannot be treated as a collision dominated, hydrodynamical fluid  and the halos show a very complex evolution
which is summarized in figure 1. In this simulation the density distribution initially is characterised by a Hernquist profile
(Hernquist 1990) which provides an excellent fit to the structure of virialized, collisionless CDM halos with a central density
cusp of $\rho \sim r^{-1}$. Note, that the velocity dispersion profile has a maximum at the inversion
radius $r_i = 0.33 r_s$ and decreases towards the center for $r < r_i$.
$r_s$ is the scale radius of the dark halo which is related to its virial radius $r_{200}$ through
the concentration $r_s = r_{200}/c$ with $c \approx 16$ for CDM halos. For the Hernquist profile
$r_i=r_s/3$.

\begin{figure}[t]
\begin{center}
\includegraphics[width=.9\textwidth]{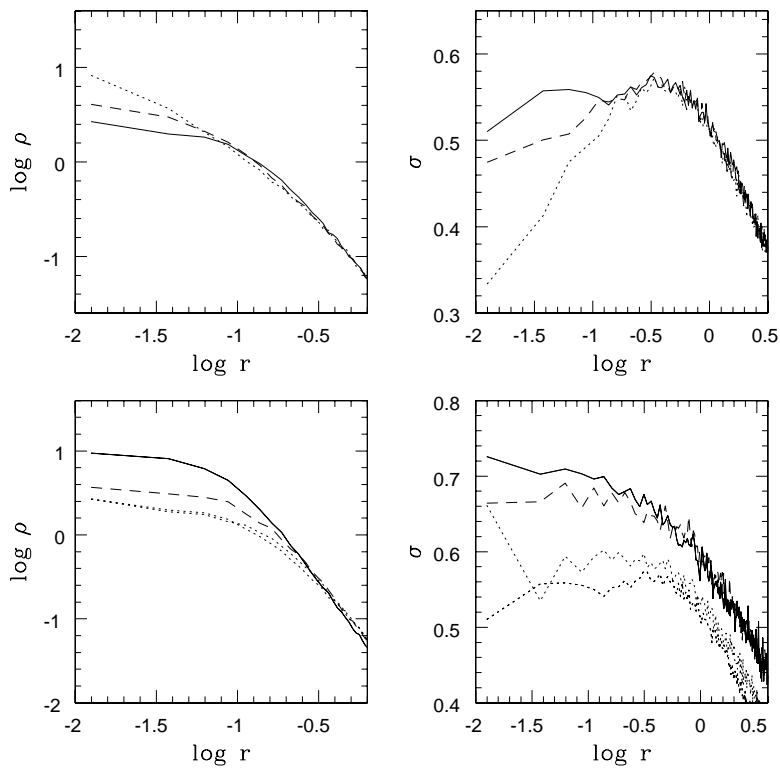}
\end{center}
\caption[]{The evolution of a weakly interacting CDM halo. The density distribution $\rho(r)$ and
the 3-dimensional velocity dispersion $\sigma(r)$ in dimensionless unts are shown. The upper
panels show the phase of core expansion with the dashed line corresponding to the initial state
and the solid line to the state of maximum expansion. The lower panels show the
subsequent epoch of core collapse.}
\label{eps1.1}
\end{figure}

Figure 1 shows the evolution of the dark matter density distribution and the velocity dispersion
profile inside the core region. As a result of energy exchange by weak interaction the 
kinetic temperature inversion leads to heat conduction inwards. The central velocity dispersion increases
with time and the core expands, resulting in a shallower density distribution. After a relaxation timescale 

\begin{equation}
\tau_{rel}=2 \left( \frac{\tau_{dyn}}{\sigma_* M_{200}/r_s^2} \right) \approx 100 \tau_{dyn}
\left( \frac{\sigma_*}{1 g cm^{-2}} \right)^{-1} \left( \frac{r_s}{kpc} \right)^{-1}
\end{equation}

\noindent an isothermal, constant density core has formed with a radius that is of order
$r_i$. Subsequently, weak interactions between the kinematically hotter core and the cooler envelope lead
to a flow of kinetic energy outwards which causes the isothermal core to contract and heat up further
due to its specifc heat, starting a core collapse phase. Detailed simulations by
Kochanek \& White (2000) have shown that on a timescale of only a few $\tau_{rel}$  an even steeper
$1/r^2$ would form again.

\section{Conclusions}
Observations indicate low-density cores in dwarf galaxies, in contrast to the predictions by
cosmological simulations. Weak self-interaction of dark matter particles could solve this problem,
leading to constant-density cores on the typical age (100 $\tau_{dyn}$) of dwarf galaxies
if $\sigma_* \geq$ 0.1 g cm$^{-2}$. Cross sections much larger than
this value would lead to core collapse and to very steep core density profiles, in contradiction
with the observations. Additional constraints are provided from the studies of galaxy clusters.
Miralda-Escud\'{e} (2000) points out that too many particle collisions would produce galaxy clusters
which are rounder than observed. More detailed calculations by Yoshida et al. (2000) demonstrate
that this constraint requires cross sections close to the minimum value that would still
alter the rotation curves in dwarf galaxies.
Combined with other arguments (Wandelt et al. 2000, Gnedin \& Ostriker 2000) this restricts the value of the
weak interaction cross-section to a narrow range around 0.1 g cm$^{-2}$.

%

\end{document}